**Title**

# Mid-infrared Chemical Imaging of Intracellular Tau Fibrils using Fluorescence-guided Computational Photothermal Microscopy

**Author list**

Jian Zhao[1, 2, *], Lulu Jiang[3], Alex Matlock[4], Yihong Xu[5], Jiabei Zhu[1], Hongbo Zhu[6], Lei Tian[1,7], Benjamin Wolozin[3], and Ji-Xin Cheng[1, 5, 7, 8, *]

**Affiliations**

[1]Department of Electrical and Computer Engineering, Boston University, Boston, MA 02215, USA
[2]The Picower Institute for Learning and Memory, Massachusetts Institute of Technology, Cambridge, Massachusetts 02142, USA
[3]Department of Pharmacology and Experimental Therapeutics, Boston University School of Medicine, Boston, MA 02118, USA
[4]Department of Mechanical Engineering, Massachusetts Institute of Technology, Cambridge, Massachusetts 02142, USA
[5]Department of Physics, Boston University, Boston, MA 02215, USA
[6]State Key Laboratory of Luminescence and Applications, Changchun Institute of Optics, Fine Mechanics and Physics, Chinese Academy of Sciences, Changchun 130033, China
[7]Department of Biomedical Engineering, Boston University, Boston, MA 02215, USA
[8]Photonics Center, Boston University, Boston, MA 02215, USA

*Corresponding authors
*E-mail: jxcheng@bu.edu (J.X.C.), jianzhao@knights.ucf.edu (J. Z.)

**Abstract**
Amyloid proteins are associated with a broad spectrum of neurodegenerative diseases. However, it remains a grand challenge to extract molecular structure information from intracellular amyloid proteins in their native cellular environment. To address this challenge, we developed a computational chemical microscope integrating 3D mid-infrared photothermal imaging with fluorescence imaging, termed Fluorescence-guided Bond-Selective Intensity Diffraction Tomography (FBS-IDT). Based on a low-cost and simple optical design, FBS-IDT enables chemical-specific volumetric imaging and 3D site-specific mid-IR fingerprint spectroscopic analysis of tau fibrils, an important type of amyloid protein aggregates, in their intracellular environment. Label-free volumetric chemical imaging of human cells with/without seeded tau fibrils is demonstrated to show the potential correlation between lipid accumulation and tau aggregate formation. Depth-resolved mid-infrared fingerprint spectroscopy is performed to reveal the protein secondary structure of the intracellular tau fibrils. 3D visualization of the β-sheet for tau fibril structure is achieved.

**Introduction**

Alzheimer's Disease (AD) affects nearly fifty million people worldwide[1]. Individuals with AD develop cognitive impairment including memory loss that eventually progresses to death. As an important type of amyloid protein aggregates, tau aggregates are a pathological hallmark of AD and related neurodegenerative disorders[2,3]. The mechanism of tau aggregate formation and the pathways underlying tau-induced diseases are still not well understood. Characterizing the geometrical morphologies and chemical structures of tau aggregates in their cellular native states can provide valuable structural and functional information useful for uncovering the mechanisms of tauopathies and for developing therapeutics. To this end, there are several challenges to overcome. First, the tau protein aggregates are predominantly intracellular, and their formation is sensitive to molecular environments[2,4,5]. It is, therefore, highly desired to perform non-invasive and intracellular characterizations, which require sub-cellular resolutions and compatibilities with intracellular fluid environments. Second, aggregates formed by different isoforms and conformers of tau exhibit remarkable structural diversity at the atomic level associated with distinct tauopathies[5-7]. The same protein can also form distinct structures with respect to both morphologies and molecular arrangements[5]. In addition, β sheets constitute the main secondary protein structures of tau aggregates and other types of amyloid protein aggregates[8,9]. Therefore, it is preferable to develop methods that can resolve 3D morphological and chemical heterogeneities and are sensitive to β-sheet secondary structure. Third, the interplay between amyloid protein aggregates, including tau aggregates, and lipid droplets might play an essential role in the formation of tau aggregations but is not fully understood[10-13], requiring a method that enables the characterizations of lipids and protein simultaneously.

To tackle these challenges, various techniques have been applied to characterize tau aggregates and other amyloid protein aggregates. Structural biology tools[14,15], including X-ray crystallography, small-angle X-ray scattering, nuclear magnetic resonance spectroscopy, and Cryo-Electron Microscopy (Cryo-EM), have played pivotal roles in determining atomic-level structures of amyloid fibrils[16-19], α-synuclein oligomers[20], tau protein[21,22], and tau filaments[23,24]. However, these methods require purified protein samples with complicated sample preparations[14,25-27], making it challenging to directly analyze intracellular tau aggregates in cells. In addition, the related instrumentations and usages are expensive[28] (e.g. ~$7 million for buying a Cryo-EM and ~$10,000 operational cost per day), imposing additional restrictions. The spectroscopy method, Circular Dichroism (CD) spectroscopy, can provide rapid secondary protein structure quantifications for samples in dilute solutions[20,29,30]. Yet, CD spectroscopy lacks high predictive accuracy for β-sheet secondary structures[29]. More importantly, the spectroscopic method usually lacks site-specific information, leading to difficulties in attributing spectral features to the structural heterogeneity of protein aggregates. Imaging methods, such as Positron Emission Tomography (PET) and fluorescence imaging, have also been applied to investigate neurodegenerative diseases. PET enables the brain-wide amyloid-β contents imaging with radiotracer but has a low spatial resolution (~2 mm)[31-33]. Fluorescence imaging is popular for high-resolution and high-fidelity intracellular imaging of tau aggregates but lacks the capabilities of quantifying secondary protein structures[4,34,35]. Overall, the abovementioned solutions are fundamentally limited by their inability to perform cost-effective, non-invasive 3D imaging and analysis of amyloid proteins' morphology and chemical structure in the cellular fluid.

Vibrational spectroscopic imaging, including Raman and Infrared (IR)-based methods, can potentially circumvent the abovementioned technique barriers based on cost-effective solutions for imaging amyloid protein aggregates in their native states. For Raman-based solutions, stimulated Raman scattering microscopy and tip-enhanced Raman spectroscopic imaging have been applied to investigate tau fibrils, amyloid plaques, and protein aggregates related to polyglutamine diseases[36-41]. However, Raman scattering is a weak process with an extremely small cross-section of ~$10^{-30}$ to ~ $10^{-28}$ cm$^2$. Due to this limitation, Raman imaging systems are based on point-scanning configurations with tightly focused high-power laser beams, resulting in low imaging speed, weak signals in the amide bands, and a high potential for photodamage[42]. In addition, Raman has difficulties in discriminating parallel β-sheet protein from anti-parallel β-sheet protein without ambiguities, while β-sheet protein constitutes the most important secondary structure of amyloid protein aggregates[8,9]. In comparison, IR-based spectroscopic imaging solutions are more suitable for detecting amyloid protein aggregates, such as tau fibrils and oligomers. First, IR absorption offers a cross-section of ~$10^{-18}$ cm$^2$ that is ten orders of magnitude larger than Raman scattering[43]. An important advantage of IR lies in its exceptional sensitivity to changes in the chemical bond strength as well as variations in β-sheet protein secondary structure in amide bands since a change of 0.02% can be easily detected[9,44]. Therefore, IR-based spectroscopic imaging methods feature higher sensitivity, greater accuracy and are free from a requirement of tightly-focused beam designs with a resulting reduction in the risk of photodamage while detecting intracellular tau protein aggregates. Conventional IR spectroscopy methods can provide sensitive and accurate protein secondary structure analysis but lack high spatial resolution and site-specific information[45,46]. Atomic Force Microscopy IR (AFM-IR) spectroscopy overcomes issues arising from low spatial resolution but is limited to 2D and requires dry samples, which could alter the structure of the tau aggregates due to changes in the native intracellular fluid environment[47,48].

The emerging Mid-IR Photothermal (MIP) microscopy[49], also called Optical Photothermal IR (OPTIR) microscopy, opens a new avenue to detect amyloid protein aggregates. MIP microscopy inherits the advantages of IR spectroscopy while enabling hyperspectral 2D/3D chemical cell imaging with diffraction-limited resolution in the visible band, and being compatible with both point-scanning and widefield imaging configurations[50-58]. Prior studies show that MIP microscopy enables 2D imaging and mid-IR fingerprint spectra extractions on polymorphic amyloid aggregates in neurons using point-scanning systems[59]. Despite this significant progress, demonstrated MIP-based characterizations of amyloid proteins still cannot uncover the 3D heterogeneity of amyloid protein aggregates using 2D imaging data and averaged spectra due to the system design. 3D MIP imaging of normal intracellular proteins is achieved by Integrating infrared excitation with optical diffraction tomography[55] and intensity diffraction tomography[58]. Yet, 3D chemical imaging of intracellular amyloid protein aggregates has not been reported. Without additional guidance, these volumetric imaging and spectroscopic analysis methods of amyloid protein aggregates suffer because the weak amyloid protein signal tends to be overwhelmed by background signals from other proteins and the immersion medium.

In this work, we present a fluorescence-guided computational MIP microscope for 3D bond-selective imaging of intracellular amyloid protein aggregates (tau fibrils) as well as lipid content in the fluid environment of a cell. Our scheme can differentiate amyloid protein aggregate signatures from the background protein signals, extract 3D site-specific mid-IR fingerprints spectra, and provide high-resolution 3D visualizations of aggregate secondary structure. Our method, termed Fluorescence-guided Bond-Selective Intensity Diffraction Tomography (FBS-IDT), integrates a simple 2D fluorescence imaging modality within a chemical intensity diffraction tomography framework. FBS-IDT uses 2D fluorescence imaging as the guide star for differentiating the amyloid protein aggregates from other proteins within a large Field Of View (FOV). Then, FBS-IDT performs time-gated pump-probe hyperspectral detection to capture 3D Refractive Index (RI) variations per mid-IR wave number to generate both 3D chemical imaging results and site-specific depth-resolved mid-IR spectra. Based on protein secondary structure analysis, 3D visualization of the secondary structure compositions can be realized. Notably, FBS-IDT is a cost-effective table-top microscope developed from commercially available components, which fit most laboratories' routine usage. Compared to various state-of-the-art techniques, FBS-IDT resolves the difficulties of 3D chemical imaging of amyloid protein aggregates and their secondary protein structure in their native intracellular fluid environment for the first time. In the following sections, we first illustrate FBS-IDT's principle and system design. Then, we demonstrate high-fidelity 3D chemical imaging of intracellular tau fibrils and lipid contents and showcase their potential correlations at the single-cell level. Next, we show that depth-resolved mid-IR fingerprint spectra can be extracted from tau fibrils and reveal the chemical structure differences between tau fibrils and diffusive tau protein. Finally, we demonstrate depth-resolved protein's secondary structure analysis and 3D visualization of tau fibrils' β sheets structure.

**Results**
**FBS-IDT principle, workflow, and instrumentation**
FBS-IDT integrates the 3D label-free chemical imaging modality with a simple 2D single-photon fluorescence imaging modality. The chemical imaging modality utilizes a pump-probe MIP tomography imaging scheme to provide 3D chemical imaging information with high temporal and spatial resolution as well as site-specific mid-IR fingerprint spectra. Within this framework, the mid-IR pump laser induces chemical-specific volumetric MIP effects, and the visible probe laser images the induced 3D RI variations. Meanwhile, the 2D fluorescence imaging modality can extract the target amyloid protein aggregates from background protein signals. This 2D image can work as the guide star for the chemical imaging modality to perform the site-specific mid-IR protein secondary structure analysis, which can determine the spectral locations and areas of secondary structures. Finally, 3D visualization of protein secondary structures for amyloid protein aggregates can be achieved using the spectral ratio map.

  The principle of 3D label-free chemical quantitative phase imaging is illustrated in **Figure 1a-d**. FBS-IDT relies on the MIP effects from the mid-IR fingerprint region, where each absorption peak corresponds to a unique molecular vibrational bond[60], to generate the molecular-specific 3D RI maps. As shown in **Figure 1a**, a pulsed and tunable mid-IR pump beam is incident on the sample and absorbed by the particular chemical compositions. This chemical-specific absorption changes the temperature and causes local sample expansions that modify the local RI distributions[61]. Meanwhile, each pulsed visible probe beam from the customized laser ring array is synchronized with the mid-IR laser pulse to image the MIP-induced RI variations. This pump-probe detection is a transient process in that the local heat can dissipate within a few microseconds to tens of microseconds[61,62]. Here, the 3D RI sample map is reconstructed using the IDT method (Methods). The IDT method implements a physics model that relates the sample's properties to the scattering information recorded by the 2D intensity images based on the first-Born approximation[63-65]. Under this approximation, the scattered field from each point within the object space is independent and allows the 3D object to be treated as an axially discretized set of decoupled 2D slices. This discretization enables slice-wise 3D reconstruction of the object's RI map using deconvolution. As shown in **Figure 1b**, the 2D intensity image captured under each oblique probe illumination contains the cross-interference information that can be mapped into the 3D frequency domain. The object's 3D RI map can be reconstructed by transforming the synthesized Ewald's sphere back to the spatial domain using the data from 16 different probe illuminations[65]. 3D RI map reconstructions are performed for both "Cold" and "Hot" states (**Figure 1c**). The "Cold" and "Hot" states are created by the high-speed modulation of the pump mid-IR laser between the "Off" and the "On" states, where the chemical-specific MIP-induced RI variations are absent or present. The chemical-specific 3D RI variation map under a particular mid-IR wavenumber is recovered by the subtraction between "Hot" and "Cold" 3D RI maps. Repeating the chemical RI map extractions for the mid-IR fingerprint region can generate the 4D hyperspectral chemical images that contain 3D site-specific mid-IR spectra information (**Figure 1d**).

Fluorescence-guided amyloid protein aggregate chemical information extractions are illustrated in **Figure 1e-g**. First, the FBS-IDT is switched to fluorescence imaging mode by turning on the 488 nm excitation laser beam and adding spectral filters (**Figure 1e**). The excitation laser beam illuminates the sample labeled by Green Fluorescent Protein (GFP). A 2D single-photon fluorescence emission intensity image is recorded on the same camera. This 2D intensity image serves as a guide star to differentiate the boundary between the normal protein and the amyloid protein aggregates. Under the guidance of the 2D fluorescence image, we can extract the site-specific mid-IR fingerprint spectra from the target areas in the 4D hyperspectral chemical images (**Figure 1f**). A peak deconvolution of amide I bands in those spectra enables the protein's secondary structure analysis[46,60] (Methods), which resolves and quantifies the positions and areas of different protein bands, such as α-helical structures and β-sheet structures (**Figure 1f**). Based on the analysis in Figure 1f, we can generate the mid-IR spectral ratio map using two 3D chemical images corresponding to selected secondary structure positions (**Figure 1g**). In this work, we mainly use this ratio map method[59,66,67] to visualize 3D β-sheet structure distribution. Overall, FBS-IDT enables the characterization of intracellular chemical compositions in cellular fluid environments, including the 4D hyperspectral chemical imaging information, the 3D site-specific mid-IR fingerprint spectroscopy, and the 3D visualization of protein secondary structure. The abovementioned information fully characterizes the intracellular amyloid protein aggregates and provides 3D chemical information for other important organelles, such as lipid droplets, correlated to protein aggregates formation.

The FBS-IDT's system design is demonstrated in **Figure 2**. More details about the hardware can be found in Methods. FBS-IDT is based on a simple brightfield transmission microscope constructed using a microscope objective, a tube lens, and a CMOS camera (**Figure 2a-b**). The unique add-on elements are the customized 450 nm laser ring array containing 16 low-cost diode lasers, the tunable mid-IR Quantum-Cascade Laser (QCL), and the 488 nm excitation laser. The mid-IR laser beam and the 488 nm excitation laser beam share the same beamline. The 450 nm laser, the mid-IR laser, and the excitation laser work as the probe beam, the pump beam, and the fluorescence excitation beam, respectively. The illumination angle per probe laser beam matches the objective's Numerical Aperture (NA), maximizing the imaging system's spatial frequency coverage. This spatial frequency enhancement expands the accessible bandwidth following the synthetic aperture principles[68] and enables the diffraction-limited resolution (~350 nm laterally, ~1.1 μm axially) equivalent to incoherent imaging systems[58]. For the MIP-induced RI variations, we use an off-axis gold parabolic mirror to loosely focus the mid-IR pump beam on the sample with a beam spot size of ≈63 μm Full width at half maximum (FWHM). This beam spot size is sufficiently large for widefield single-cell imaging. During data acquisition, FBS-IDT is first operated under the fluorescence mode using the 488 nm excitation laser beam (**Figure 2a**). After obtaining the 2D guide-star image, the FBS-IDT turns on the pump-probe detection beams and performs 3D chemical quantitative phase imaging (**Figure 2b**). Since the 3D chemical quantitative phase imaging detects the transient RI fluctuation within the microsecond time scale, time synchronization is essential to capture high Signal-to-Noise Ratio (SNR) images. To this end, both the probe laser and the mid-IR laser are modulated under pulse modes (~10 kHz repetition rate and ~1 μs pulse duration) for transient pump-probe photothermal detection (**Figure 2c**). An additional 50 Hz modulation is applied to the mid-IR laser to create "Hot" and "Cold" states since the CMOS camera is running at 100 Hz. Based on this design, the 3D chemical quantitative phase imaging speed can reach up to ~6 Hz[58]. Overall, FBS-IDT features a simple and cost-effective system design and enables intracellular volumetric high-resolution, high-speed chemical imaging and site-specific spectroscopic analysis of amyloid protein aggregates in intracellular fluid environments.

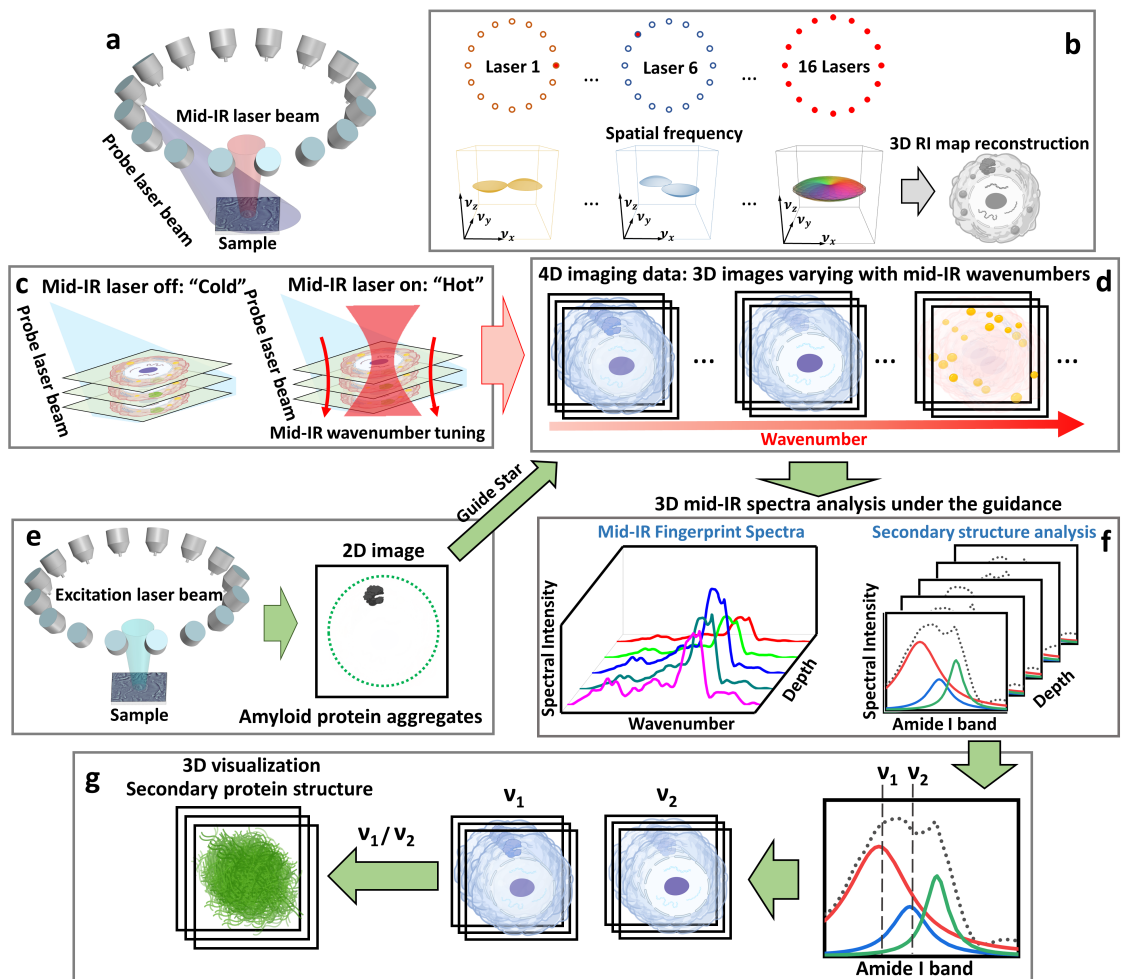

**Figure 1 FBS-IDT principle and workflow. a** Pump-probe 3D chemical imaging scheme. Each oblique pulsed probe beam (~450 nm) from a ring laser array illuminates the sample sequentially. A loosely focused pulsed mid-IR laser pump beam heats the sample periodically. **b** 3D RI map reconstruction scheme. Intensity imaging data from each probe beam illumination are mapped into the frequency domain. 3D RI map can be reconstructed by the inverse Fourier transform of the synthesized frequency domain information from all 16 probe beam detections. **c** "Cold" state: imaging without mid-IR pump beam illumination; "Hot" state: imaging with mid-IR pump beam illumination. **d** 4D hyperspectral chemical imaging. 3D chemical maps under different mid-IR wavenumbers are obtained in two steps: 1) subtracting "Hot" 3D RI maps from the "Cold" 3D RI maps under a particular wavenumber; 2) tuning the wavenumber to obtain 3D maps for different chemical compounds of interest. **e** Single-photon 2D fluorescence intensity imaging. To obtain the 2D guide star, both probe and pump beams are turned off while an excitation laser beam illuminates the sample. The 2D fluorescence images highlight the boundary of the amyloid protein aggregates. **f** Depth-resolved mid-IR fingerprint spectra generation and related protein secondary structure spectroscopic analysis. The depth-resolved mid-IR spectra are extracted from the 4D hyperspectral chemical imaging data under the guidance of the 2D image in "e". Secondary structures are analyzed by the deconvolution of the mid-IR amide I band. **g** 3D visualization of protein secondary structure for the amyloid protein aggregates. Based on the spectroscopic analysis results in **f**, spectral positions for specific secondary protein structures are selected. 3D visualization of the secondary protein structures is obtained by extracting the mid-IR spectral ratio map between two 3D chemical images. Cell and protein aggregate icons in "b-e" and "g" are created and adapted from[69].

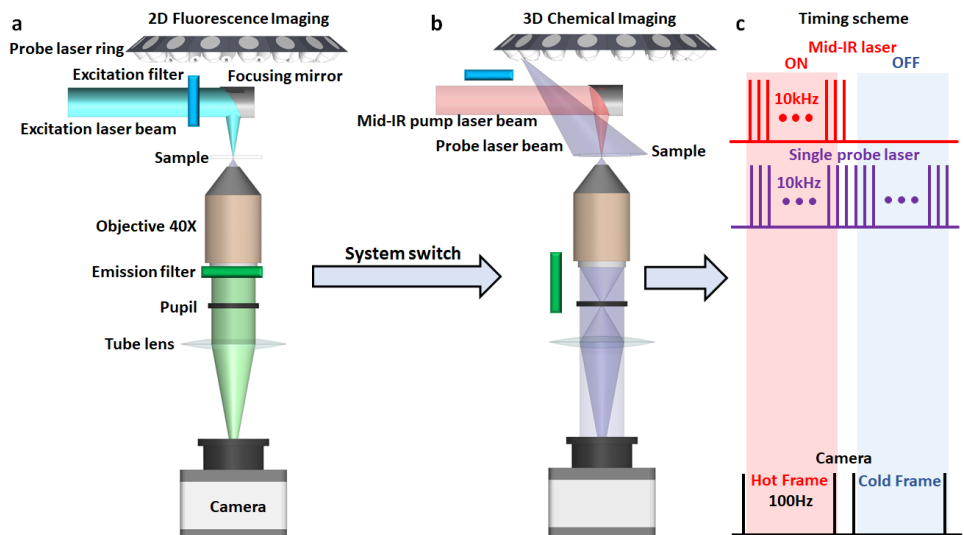

**Figure 2. FBS-IDT instrumentation.** FBS-IDT is based on a widefield transmission microscope consisting of a 40x microscope objective, a tube lens, a CMOS camera, and add-on modalities that include a probe laser ring, a mid-IR pump laser, and an excitation laser. **a** 2D single-photon fluorescence intensity imaging mode. Under this mode, both the excitation filter and the emission filter are turned on. The 488-nm excitation laser beam is loosely focused on the sample using an off-axis parabolic mirror. **b** 3D chemical imaging mode. The oblique illumination of each probe beam matches the objective's NA. The pump beam shares the same beam path with the 488-nm excitation laser beam. Under this mode, both pump and probe lasers are turned on and illuminating the sample while the excitation laser is switched off. **c** Time synchronization scheme for 3D chemical imaging. Both probe laser and mid-IR laser are running at 10 kHz with a pulse duration of ~ 1 µs. Each probe pulse is synchronized with a mid-IR laser pulse with a time delay of ~ 0.5 µs. An additional 50 Hz on/off duty-cycle modulation is imposed on the mid-IR laser to generate "Hot" and "Cold" frames on the camera. The camera is running with a frame rate of 100 Hz.

### 3D bond-selective imaging of intracellular lipids and proteins

Lipid homeostasis plays a vital role in maintaining normal neuronal functions. Dysregulation of lipid metabolism has been observed to correlate with neurodegenerative diseases[10,13]. Previous imaging research confirmed the co-localization of lipid compositions with amyloid protein deposits in tissues[70-72]. Despite this progress, the role of lipids in neurodegenerative diseases is still not clear. It requires imaging tools to investigate the interplay between lipids and amyloid protein aggregates. To this end, 3D chemical imaging is highly desired for visualizing different intracellular chemical compositions in a fluid environment. In **Figure 3**, we demonstrate label-free hyperspectral 3D chemical imaging on fixed human epithelial cells (Tau RD P301S FRET Biosensor cells). The cells in the experimental group were cultured with seeded tau fibril fractions, while the cells in the control group were cultured without tau fibril seeding (Methods). All the cell samples were fixed, washed, and immersed in $D_2O$ Phosphate-Buffered Saline (PBS) between two pieces of 0.2 mm-thick Raman-grade $CaF_2$ glasses. $D_2O$-based PBS demonstrates a relatively flat spectral response and minimizes the overlap between the amide I protein band and the O-H bending vibration from water[73]. Cells in both the experimental and control groups were imaged under the same mid-IR laser illuminations. **Figure 3a-d** (see Figure S1 in the Supplementary Information for more data) demonstrate depth-resolved reconstructions of the cell with tau fibrils under four different conditions: a) RI map without mid-IR laser illumination; b) protein absorption map in the amide I band; c) lipid absorption map; d) RI variation map at the cell-silent window (1800 cm$^{-1}$ – 2700 cm$^{-1}$) where minimal RI changes are expected. Based on the above imaging results, the variation of cellular features along different axis positions can be clearly observed. The chemical-specific cellular features only appear under the corresponding mid-IR fingerprint wavenumbers, resulting in the distinct morphology between the protein and the lipid maps. The chemical images from the cell-silent window confirm that the imaging contrast originates from MIP-induced absorption.

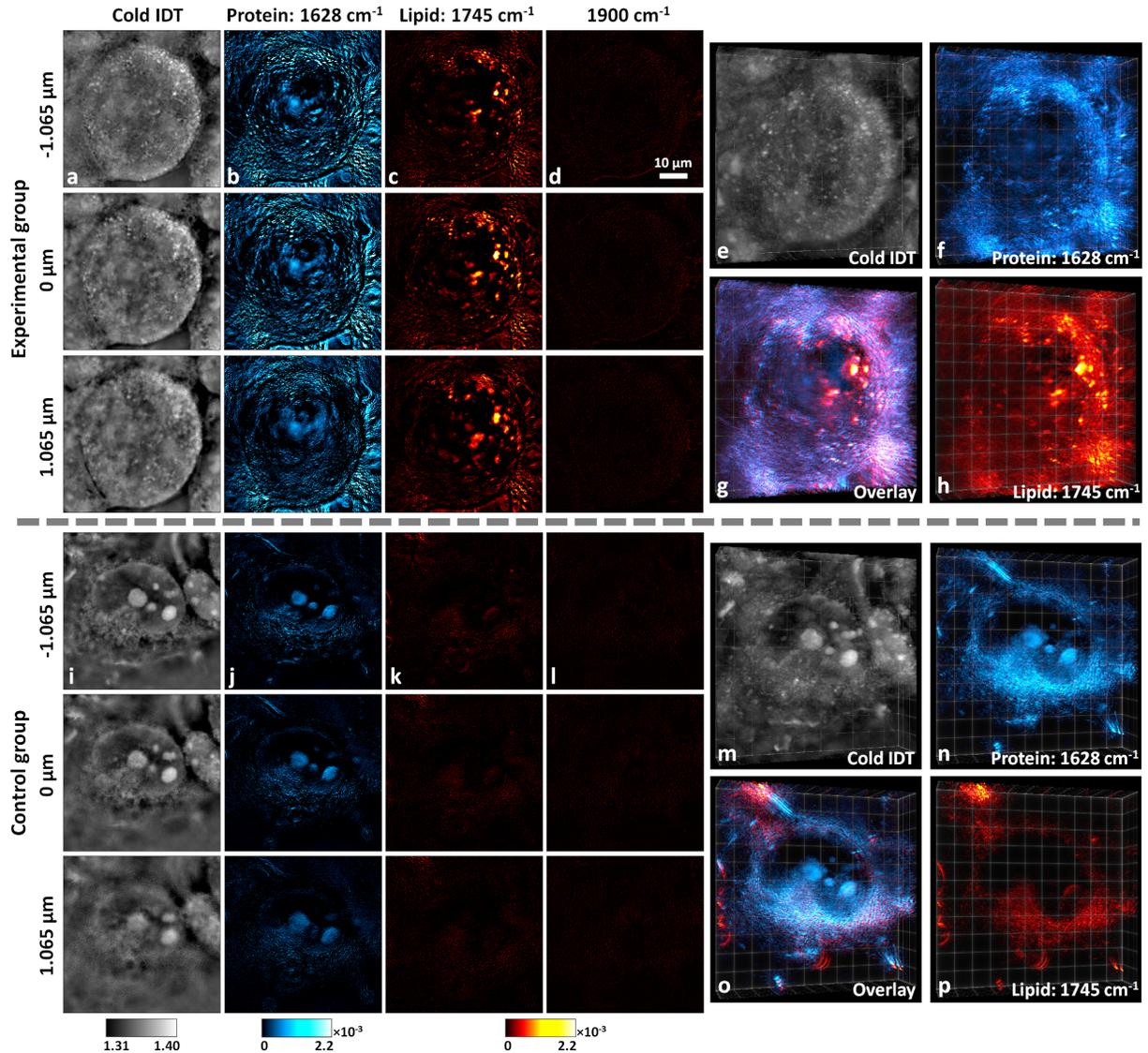

**Figure 3. FBS-IDT chemical imaging on human epithelial cells (Tau RD P301S FRET Biosensor cells). a-h** 3D chemical imaging of the experimental group with fibrillar tau aggregates. **i-p** 3D chemical imaging of the control group without fibrillar tau aggregates. The fixed cells for both groups are immersed in $D_2O$ PBS. The scale bar in "d" is applicable to "a-d" and "i-l". **a, i** Depth-resolved cold imaging results. **b, j** Protein imaging results with 1628cm$^{-1}$ mid-IR wavenumber. **c, k** Lipid imaging results with 1745cm$^{-1}$ mid-IR wavenumber. **d, l** Off-resonance imaging results with 1900 cm$^{-1}$ mid-IR wavenumber. Most biochemical compounds have significantly weak or no absorptions at this cell-silent wavenumber. **e, f, h** 3D rendering of the cells in the experimental group shown in "a", "b", and "c". **g** Overlay chemical imaging results of "f" and "h". **m, n, p** 3D rendering of the cells in the control group shown in "i", "j", and "k". **o** Overlay chemical imaging results of "n" and "p". Extended data can be found in Figure S1 in the Supplementary Information.

Here, the experimental group's protein-based signal (**Figure 3b**) is distributed throughout the cell, making it hard to locate the tau fibrils. Besides the protein map, the lipid map (**Figure 3c**) of the experimental group demonstrates high concentrations of lipid compositions, which might correlate to the tau fibrils formation in the cell[13,72]. For the control group without tau fibrils, chemical images (**Figure 3i-l**) are acquired under the same conditions as the cell in the experimental group.

We can clearly observe the chemical-specific depth-resolved cellular structures based on the control group's chemical imaging results. Compared to the experimental group, the significant difference is mainly manifested in the lipid map, where the control group has a low concentration of lipid contents. This stark difference in lipid concentrations between the control and experimental groups matches previous observations[70-72], which might correlate to the tau fibrils formation. Besides the depth-resolved 2D demonstrations, we further perform 3D rendering of the chemical imaging results for the experimental group (**Figure 3e-h**, Supplementary Movie 1) and the control group (**Figure 3m-p**, Supplementary Movie 2). These 3D chemical images and movies clearly illustrate the volumetric distribution of the chemical-specific cellular structures. The overlay 3D images simultaneously show the protein and lipid volumetric distribution, providing rich chemical and spatial correlation information. Although proteins with different secondary structure compositions are supposed to show varying MIP absorption, it is still difficult to directly extract this small variation. This gap can be filled when the 2D fluorescence guidance comes into play, as shown below.

**Depth-resolved mid-IR fingerprint spectra of tau fibrils**

Amyloid protein aggregates exhibit highly heterogeneous chemical structures and morphologies[5,7,74]. The chemical and morphological variations show different levels of neurotoxicity and are associated with different stages of AD[74-76]. This heterogeneity is site-specific and sensitive to the cellular environment since the formation of the aggregate structure is subject to post-translational modifications and cofactors[5,7]. The chemical, morphological and functional heterogeneity in AD is poorly understood, especially for the prefibrillar and early-stage aggregates. FBS-IDT enables 3D chemical imaging and 3D site-specific mid-IR spectra extraction for intracellular protein aggregates in the cellular fluid environment. Here, we demonstrate FBS-IDT's capabilities of depth-resolved mid-IR fingerprint spectra extractions (**Figure 4**).

We first performed single-photon 2D intensity fluorescence imaging on cells with/without tau fibrils (**Figure 4b, e**). As expected, the experimental group shows strong GFP-emission signals from labeled fibrillar tau aggregates, while no tau aggregates are observed in the control group. Fluorescence imaging results confirm the existence of tau fibrils in the experimental group and provide a guide star for extracting mid-IR spectra for tau fibrils. To extract depth-resolved mid-IR spectra, we first obtain hyperspectral 3D chemical images (4D data) of the cell samples for both experimental and control groups, covering the mid-IR fingerprint region. It is well known that the major absorption peak of the β-sheet structure is around 1630 $cm^{-1}$ in the amide I band[8,44]. We thereby selected the areas of interest with high SNR for extracting mid-IR spectra from the tau fibrils using a protein map at 1628 $cm^{-1}$ under fluorescence guidance (**Figure 4c**). We further generated the depth-resolved mid-IR fingerprint spectra from the selected areas by averaging voxels within each axial plane. The depth-resolved mid-IR fingerprint spectra for tau fibrils are shown in **Figure 4g, i, k**. As a comparison, we performed similar operations for the cell sample in the control group. The spectra plots for the control group are demonstrated in **Figure 4h, j, l**. For both the experimental group and the control group spectra, the signature amide-I absorption peak (~1650 $cm^{-1}$)and two amide-II bands (~1450 $cm^{-1}$ and ~1550 $cm^{-1}$) can be observed. Here, amide I vibration depends on the secondary structure of the polypeptide backbone and is hardly affected by the sidechain[44]. It is highly sensitive to secondary structure variations and is commonly applied to protein secondary structure analysis[44]. Therefore, we focus on the amide I band for analyzing the cell samples. As illustrated in **Figure 4g, i, k** (see Figure S2 in the Supplementary Information for more data), the amide I bands' shapes significantly vary with the axial depths for the experimental group, which shows the heterogeneity of the tau fibrils. In contrast, the amide I bands of the control group in **Figure 4h, j, l** (see Figure S2 in the Supplementary Information for more data) demonstrate slight variations, consistent with the relatively uniform distribution of the diffusive protein.

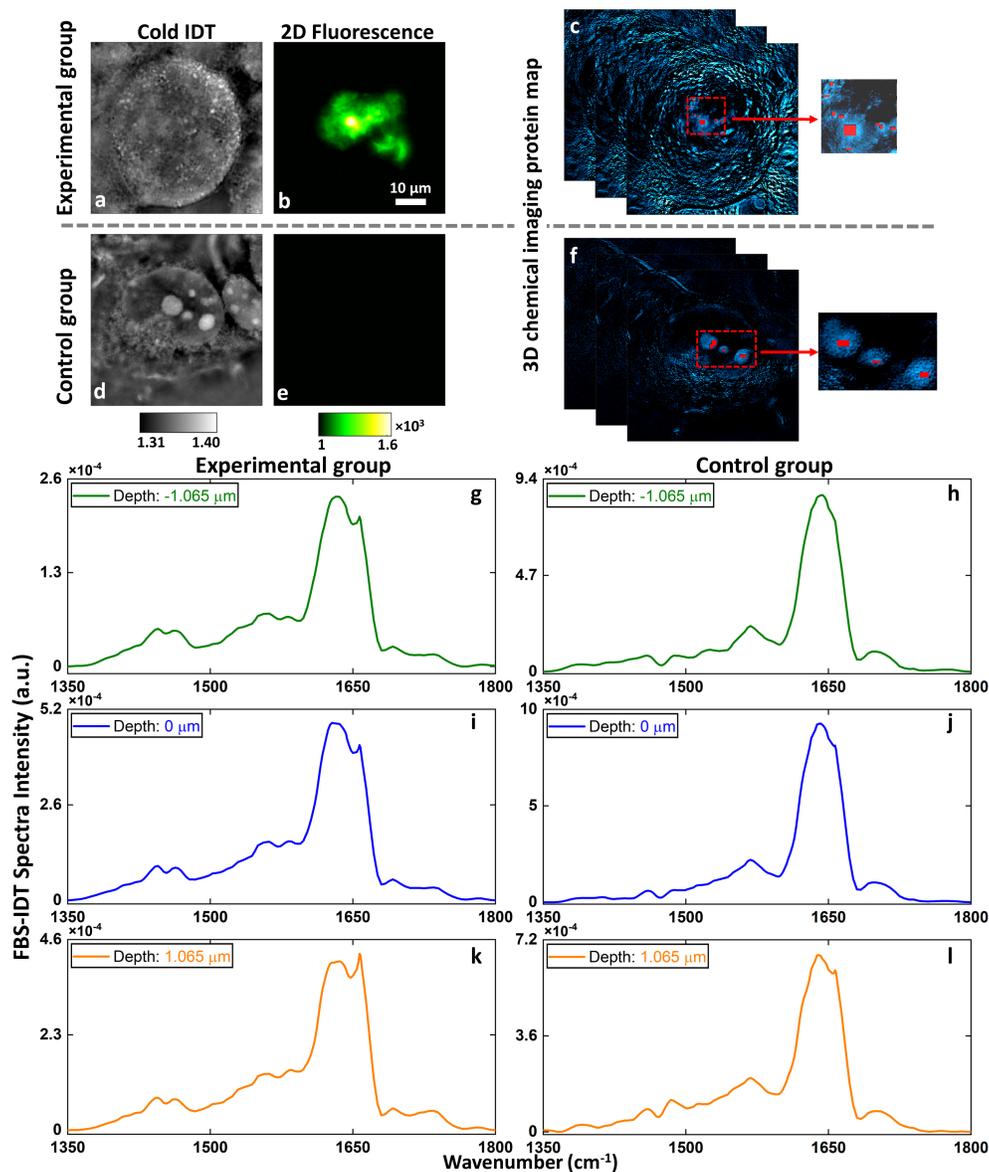

**Figure 4. Fluorescence-guided depth-resolved mid-IR fingerprint spectra of tau fibrils. a** Cold IDT image at -1.065 µm depth for the cell with tau fibrils. **b** 2D single-photon fluorescence intensity image of the cell shown in "a". **c** Select areas within the tau fibrils from the 3D chemical protein map at 1628 cm$^{-1}$. The selection is guided by the 2D image in "b" and highlighted in red. **d** Cold IDT image at -1.065 µm depth for the cell without tau fibrils. **e** 2D single-photon fluorescence intensity image of the cell shown in "d". **f** Select areas that show protein signals from the 3D chemical protein map at 1628 cm$^{-1}$. The selection is highlighted in red. **g-k** Depth-resolved mid-IR spectra extracted from the areas shown in "c". **h-l** Depth-resolved mid-IR spectra extracted from the areas shown in "f". The axial depth value of each spectrum is shown in the top left of each plot for "g" to "l". Extended data can be found in Figure S2 in the Supplementary Information.

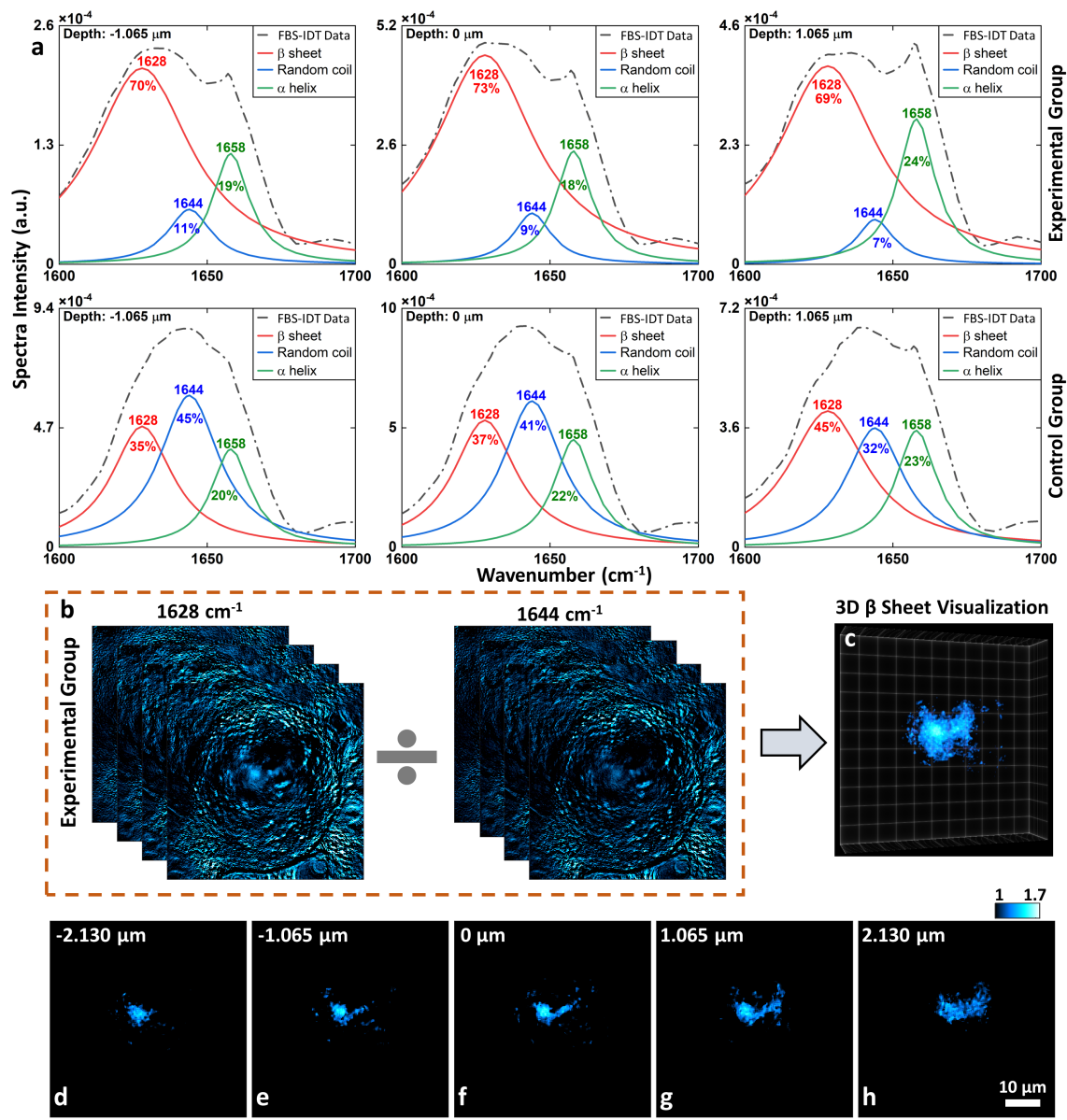

**Figure 5. Protein secondary structure analysis and 3D visualization of β-sheet structure. a** Protein secondary structure analysis based on amide I bands for cells in the experimental and the control groups. The amide I band spectra are extracted from the mid-IR fingerprint spectra shown in Figure 4. Three main protein secondary structures, the α helix, β sheet, and random coil, are quantified using the deconvolution method. The percentage and the peak positions for each secondary structure are indicated in each plot. **b** 3D protein maps of the cell with tau fibrils in the amide I band. These two images are selected according to the spectral positions of the β sheet and random coil. **c** 3D visualization of β sheet structure based on mid-IR spectral ratio map of two 3D images in "b". **d-h** Depth-resolved demonstration of the 3D rendering image in "c". The axial depth value is indicated in each image. The scale bar in "h" is applicable from "d" to "h". Extended data can be found in Figure S3 in the Supplementary Information.

**Protein secondary structure analysis and 3D visualization of β-sheet structure**

Based on the depth-resolved mid-IR fingerprint spectra shown in **Figure 4**, we further performed the protein secondary structure analysis for the experimental and the control groups. As shown in **Figure 5a** (see Figure S3 in the Supplementary Information for more data), we performed the Fourier self-deconvolution[46,60] on amide-I band spectra to quantify the main protein secondary structure compositions, such as β sheet, α helix, and random coil (Methods). The protein secondary structure analysis of the experimental group confirms that the β sheet is the main secondary structure of the tau fibrils, which is consistent with previous observations by Cryo-EM before[23,24]. The variations of the experimental group's secondary structure compositions further show the tau fibrils' chemical structure heterogeneity. In comparison, the analyses of the control group that consists of diffusive proteins demonstrate much lower percentages of β sheet, indicating relatively homogeneous chemical structures. Overall, the depth-resolved mid-IR spectra and related protein secondary structure analyses validate the existence, the chemical compositions, and the heterogeneity of the intracellular tau fibrils, which are consistent with our fluorescence imaging and known facts.

To visualize 3D protein secondary structure, we further selected two 3D protein maps from the experimental group's 4D hyperspectral data (**Figure 5b**). We obtain the mid-IR spectral ratio map by dividing the 1628 cm$^{-1}$ map by the 1644 cm$^{-1}$ map[59,66,67]. After denoising processing, 3D visualization of β-sheet structure in tau fibrils was achieved (Methods). 3D rendering of the β-sheet structure is shown in **Figure 5c** and Supplementary Movie 3. We further demonstrate 2D cross-sections along different axial depth positions within the β-sheet 3D reconstruction in **Figure 5d-h**. Although Cryo-EM enables high-resolution 3D characterization of tau aggregates, it is limited to purified protein extracted from cells with complicated sample preparations. Here, our FBS-IDT neither requires purified protein nor imposes any additional requirements on the protein aggregate sample preparations. It can directly perform 3D chemical non-invasive imaging of the intracellular protein aggregates and their secondary structure in the cell fluid with an optical approach.

**Discussion**

Based on simple and cost-effective instrumentation, FBS-IDT enables chemical imaging on lipid droplets and protein, protein aggregates characterizations inside the cellular fluid, and 3D site-specific mid-IR spectroscopic analysis, which are not demonstrated by other state-of-the-art methods. Specifically, FBS-IDT realizes label-free hyperspectral 3D quantitative chemical imaging of intracellular tau fibrils and lipid droplets in the cellular fluid with a high speed and a high resolution. Based on the hyperspectral imaging capabilities, FBS-IDT further achieves 3D site-specific mid-IR spectroscopy analysis on the tau fibrils and fluorescence-guided 3D visualization of this aggregate's protein secondary structure for the first time.

    FBS-IDT's superior performance is rooted in the innovative instrumentation design by integrating fluorescence and MIP imaging into the computational imaging framework. First, FBS-IDT's pump-probe pulsed IDT imaging design can directly correlate each voxel in the 3D object to its unique mid-IR spectra with a high resolution. With this design, FBS-IDT efficiently recovers the 3D RI information from the intensity images and generates a stack of 3D chemical images using biological objects' RI variations. Since each 3D image correlates with the unique MIP effect under a particular mid-IR wavenumber, extracting the voxel's value from the same spatial position of all the 3D images can recover a site-specific mid-IR fingerprint spectrum. Unlike FBS-IDT, most conventional spectroscopy approaches, such as Raman spectroscopy, CD spectroscopy, and FTIR spectroscopy, merely provide spectra averaged over the sample volumes without 3D spectroscopic analysis capabilities. Recently emerging AFM-IR spectroscopy can extract IR spectra from specific sites but is limited to 2D plane[47,48]. In addition to the above advantages, FBS-IDT can convert the 3D site-specific mid-IR spectra into high-resolution images, circumventing the low-resolution restrictions of conventional IR-based spectroscopic imaging methods, such as FTIR micro-spectroscopy[77]. This is achieved by separating the detection process from the IR absorption process. The 450-nm short wavelength detection guarantees high

spatial resolution. Meanwhile, the oblique illumination further enhances the resolution up to the incoherent resolution limit. Second, FBS-IDT applies a tunable, broadband QCL laser for mid-IR fingerprint spectroscopic imaging, laying the foundation for high-sensitivity bond-selective imaging and protein secondary structure analysis for amyloid protein aggregates. As reported in the previous research[44], mid-IR spectroscopy is sensitive to a small fraction (~0.02 %) of chemical bond strength change and can reach spectroscopy resolution up to ~0.2 Angstrom, which guarantees FBS-IDT's chemical sensitivity. Furthermore, due to its reliable quantification capability, the mid-IR amide I band has been commonly applied to perform protein secondary structure analysis[9,44]. Notably, it is exceptionally sensitive to β-sheet secondary structure in amyloid protein aggregates and can efficiently track protein aggregation kinetics[9,44]. Compared with mid-IR-based FBS-IDT, Raman spectroscopy requires a much higher concentration of analytes to obtain reasonable SNR, making it not commonly applied in protein secondary structure analysis[9]. Another important spectroscopy method, CD spectroscopy, is less accurate in analyzing β-sheet structure than other secondary structures[29], limiting its applications in amyloid protein. Besides the advantages of imaging and analyzing amyloid protein aggregates, FBS-IDT's mid-IR fingerprint spectroscopic imaging significantly simplifies the requirements for fluorescence tag. Unlike fluorescence-only imaging methods, FBS-IDT merely needs a single fluorescent tag for a certain type of protein since FBS-IDT can easily distinguish different biomolecules, such as lipids, DNA, and RNA, from proteins relying on label-free MIP effects. This can avoid spectral overlap issues from multiplexed fluorescence tags. Third, different from electron microscopy, FBS-IDT works in the visible and IR optical bands with a widefield configuration, compatible with fixed/living biological objects in the native cellular fluids and free from protein purification requirements. Here, both visible and mid-IR illuminations are non-toxic and harmless to most living biological entities. Also, the widefield illumination beams are either fast-diverging or loosely focused, resulting in low light intensity on the sample. Hence, the FBS-IDT's optical imaging scheme guarantees non-invasive, safe, and low photodamage biological imaging in objects' native states. Compared to FBS-IDT, Cryo-EM mainly applies to purified protein samples[23,24] with complicated sample preparations[27], hindering the investigations of the intracellular amyloid protein aggregates. Another emerging technique, AFM-IR spectroscopy imaging, does not require a purified protein sample but is limited to dry samples[47,48]. The dehydration process can alter the pathological protein aggregate since they are sensitive to the environment[5]. Fourth, the FBS-IDT is a universal and scalable imaging platform with an affordable cost and negligible daily operational expenses. FBS-IDT does not require specialized optics and can be realized by adding CW diode lasers and a mid-IR laser to a standard brightfield microscope. The cost of FBS-IDT is affordable to most laboratories compared to Cryo-EM machine's several million dollars cost and ~10k dollars daily operational fee[28].

FBS-IDT's system performance still has room for improvement. First, the fluorescence tag inevitably perturbs cellular functions and tends to bring in photobleaching issues in the current design. To realize fully label-free imaging, a deep neural network could be trained to extract the weak contrast of the target protein aggregates from the background protein signal and effectively denoise the extracted image. Second, the linear IDT reconstruction method degrades with high NA and strong scattering biological samples. The recently developed non-paraxial multiple-scattering model[78] can be deployed within the FBS-IDT framework to significantly improve FBS-IDT's resolution and high-fidelity imaging capabilities on strong scattering biological samples. Third, the high-energy nanosecond mid-IR optical parametric oscillator could replace the current low-energy QCL laser, improving the SNR and expanding the FOV.

FBS-IDT's application potential can be further explored to support tauopathy research by performing systematic imaging investigations in the future. There are many fundamental questions to be answered in tauopathy, such as the mechanism of tau aggregate generation/propagation and the role of tau oligomers in neuron toxicity[2]. To answer these questions, tau aggregation must be investigated at both single-cell and tissue levels to elucidate how tau aggregates vary in spatial distribution, chemical/morphological heterogeneity, conformation, relationship with environments, and interaction with organelles. Related biological samples can be from cultured cells/tissues or model animals. Future studies will need to focus on *in vivo* imaging to monitor the dynamics of tau and its aggregation. FBS-IDT is particularly suitable for the aforementioned application scenarios. It enables high volumetric imaging speed, non-invasive 3D spectroscopic imaging and analytic capabilities, and compatibility with biological analytes in cellular fluids. FBS-IDT will enable the comprehensive characterization of tau aggregates under numerous conditions.

In summary, using a cost-effective optical imaging system, FBS-IDT realizes 3D chemical imaging and spectroscopic analysis of intracellular lipid droplets, tau aggregate, and the secondary structure of tau aggregates in the cellular fluid. FBS-IDT enables 3D spatially-resolved mid-IR fingerprint spectra extraction and quantitative chemical imaging with sub-micrometer spatial resolution and Hz-level speed. The above spectroscopic chemical imaging does not impose additional restrictions on the biological samples, opening a new avenue for *in vivo* imaging of intracellular protein aggregates. Notably, FBS-IDT is based on a modular design developed from low-cost standard optics components, which can be easily adapted to meet different imaging needs. We envision that the FBS-IDT can significantly contribute to neurodegeneration research and a broad range of biomedical applications.

## Methods
### Instrumentation
The transmission brightfield microscope for FBS-IDT is a simple 4f imaging system built using the Thorlabs off-the-shelf components, including a microscope objective (RMS40x,0.65 NA, 40x magnification), a tube lens (TTL180A, f=180mm), a silver plane mirror, and standard 60-mm cage assembling system. The camera is an Andor CMOS camera (ZYLA-5.5-USB3-S). The 488 nm excitation laser beam for the 2D single-photon intensity fluorescence imaging is generated from a tunable femtosecond laser (Coherent, Chameleon Ultra) by frequency doubling its 976 nm wavelength. The excitation laser beam power is ~30 mW. The fluorescence imaging in FBS-IDT can also use the 488 nm beam from a regular CW diode laser. We installed switchable excitation filters (ET485/20 from Chroma) and emission filters (FF01-525/45 from Semrock and FELH0500 from Thorlabs) in the FBS-IDT beamline to switch between the fluorescence imaging mode and the chemical imaging mode. For the label-free chemical imaging, we added a ring-shaped probe laser system and a mid-IR laser to the brightfield microscope. The probe laser beam is from a customized ring laser illumination system that consists of 16 individual fiber-coupled (multimode optical fiber: 0.22 NA and 105 μm core diameter) diode lasers (wavelength: ~450 nm, average power under CW mode: ~3 W, repetition rate: up to 10kHz, pulse duration: ~0.6 μs to ~10μs). Under pulsed mode, the probe laser output power is ~30 mW based on a ~0.01 duty cycle. Each probe beam's illumination angle is adjusted to match the microscope objective's NA. The pump laser beam is from a tunable mid-IR QCL laser (Daylight solution MIRcat-2400). A gold parabolic mirror (Thorlabs, MPD01M9-M03) redirects the excitation laser beam or the mid-IR pump beam into the sample. For pump-probe chemical imaging, the pulse duration and the repetition rate are set to ~1 μs and 10 kHz for both probe and pump beam, while the Andor camera runs at a 100 Hz frame rate. The energy fluence of the probe beam on the sample area is ~0.2 pJ/μm$^2$. The mid-IR energy fluence on the sample is around 50 pJ/μm$^2$, depending on the wavenumber. Additional duty cycle control is applied to the mid-IR pump laser so that the 10 kHz pulse train is turned on and off at 50 Hz. During data acquisition, the probe laser, the pump laser, and the camera are synchronized using a pulse generator (Quantum Composers 9214). The FBS-IDT system control and data acquisitions are based on customized MATLAB codes.

**Imaging method**
For 2D single-photon intensity fluorescence imaging, FBS-IDT acquires the intensity image using the CMOS camera. For the area of interest smaller than the excitation beam size, the intensity image can be directly used as the guide star. For areas larger than the excitation beam size, FBS-IDT can scan the excitation beam to extend its FOV. The scanned images are stitched using the standard deviation projection (Fiji ImageJ, Version:1.53 c).

For label-free chemical imaging, FBS-IDT is using a similar approach developed by Zhao and Matlock et al[58]. The 3D RI map reconstruction follows the conventional IDT model published by Ling et al.[64] and the annular IDT work by Li and Matlock et al[63]. For FBS-IDT, the biological object is modeled as a 3D scattering potential within a given volume $\Omega$ as $V(\mathbf{r}, z) = k^2(4\pi)^{-1}\Delta\epsilon(\mathbf{r})$, where $r$ denotes the spatial coordinates $\langle x, y, z \rangle$, $k$ is the probe beam's wavenumber, and $\Delta\epsilon(\mathbf{r})$ is the permittivity contrast between the biological object and its immersion medium. Each oblique probe beam is treated as a plane wave $u_i(\mathbf{r}|\mathbf{v}_i)$ incident on the sample with a illumination angle defined by the lateral spatial frequency vector $\mathbf{v}_i$. Based on the first-Born approximation, the total scattering field from the biological object generated by the probe beam can be evaluated as a summation

$$u_{tot}(\mathbf{r}|\mathbf{v_i}) = u_i(\mathbf{r}|\mathbf{v_i}) + \iiint_\Omega u_i\left(\mathbf{r}'|\mathbf{v_i}\right) V\left(\mathbf{r}'\right) G\left(\mathbf{r} - \mathbf{r}'\right) d^3\mathbf{r}', \qquad (1)$$

of the incident and first-order scattered field defined by a 3D convolution with the Green's function $G(r)$. Our model assumes that the total scattered field results from a stacked set of 2D slices through the object since the scattering events from each point in the object are mutually independent. This assumption implies that the 3D RI map can be recovered from a single 2D plane if the additional propagation is included in the inverse model for recovering each axial slice. FBS-IDT relates the object's 3D scattering potential with the measured intensity images using the cross interference between the incident and the scattered field from the probe laser. The cross interference linearly encodes the scattering potential into intensity. Based on oblique probe beam illumination, the cross-interference term and its conjugate are spatially separated in the Fourier plane allowing for linear inverse scattering models under the first-Born approximation. With this separation and the complex permittivity contrast assumption ($\Delta\epsilon(\mathbf{r}, z) = \Delta\epsilon_{re}(\mathbf{r}, z) + j\Delta\epsilon_{im}(\mathbf{r}, z)$), a forward model can be developed

$$\hat{I}(x, y|\mathbf{v_i}) = \sum_m H_{re}(\mathbf{v}, m|\mathbf{v_i})\Delta\hat{\epsilon}_{re}(\mathbf{v}, m) + H_{im}(\mathbf{v}, m|\mathbf{v_i})\Delta\hat{\epsilon}_{im}(\mathbf{v}, m), \qquad (2)$$

where $\hat{}$ denotes the Fourier transform of a variable, $m$ denotes the axial slice index, and $H_{re}$ and $H_{im}$ are the transfer functions (TFs) containing the physical model. These transfer functions have the form

$$H_{re}(\mathbf{v}, m|\mathbf{v_i}) = \frac{jk^2\Delta z}{2} A(\mathbf{v_i}) P(\mathbf{v_i}) \left[ P(\mathbf{v} - \mathbf{v_i}) \frac{e^{-j[\eta(\mathbf{v}-\mathbf{v_i}) - \eta(\mathbf{v_i})]m\Delta z}}{\eta(\mathbf{v}-\mathbf{v_i})} - P(\mathbf{v} + \mathbf{v_i}) \frac{e^{j[\eta(\mathbf{v}+\mathbf{v_i}) - \eta(\mathbf{v_i})]m\Delta z}}{\eta(\mathbf{v}+\mathbf{v_i})} \right], \quad (3a)$$

$$H_{im}(\mathbf{v}, m|\mathbf{v_i}) = -\frac{k^2\Delta z}{2} A(\mathbf{v_i}) P(\mathbf{v_i}) \left[ P(\mathbf{v} - \mathbf{v_i}) \frac{e^{-j[\eta(\mathbf{v}-\mathbf{v_i}) - \eta(\mathbf{v_i})]m\Delta z}}{\eta(\mathbf{v}-\mathbf{v_i})} + P(\mathbf{v} + \mathbf{v_i}) \frac{e^{j[\eta(\mathbf{v}+\mathbf{v_i}) - \eta(\mathbf{v_i})]m\Delta z}}{\eta(\mathbf{v}+\mathbf{v_i})} \right], \quad (3b)$$

where $A(\mathbf{v_i})$ is an illumination light amplitude, $P(\mathbf{v})$ is the pupil function, $\Delta z$ is the discretized slice thickness based on the microscope's depth of field, $\eta(\mathbf{v}) = \sqrt{\lambda^{-2} - |\mathbf{v}|^2}$ is the axial spatial frequency, and $\lambda$ is the imaging wavelength. Given this linear forward model, the model inversion

can be performed by slice-wise deconvolution with Tikhonov regularization. The image reconstruction is based on customized MATLAB codes.

Based on the above method, FBS-IDT can reconstruct the 3D RI maps for the object under "Cold" and "Hot" states. The 3D chemical image is obtained by subtraction between the "Hot" from the "Cold". For 3D visualization of the protein secondary structure, two 3D chemical images of interest are selected and normalized by their corresponding mid-IR laser power. Then, the division between the two 3D chemical images is performed to generate a 3D ratio map that shows the 3D contrast of the secondary structure. Then, the 3D ratio map is further denoised using thresholding and median filtering (MATLAB function: medfilt2). The 2D fluorescence image can also delineate the boundary of tau fibrils to help reduce extra noise.

**FBS-IDT mid-IR spectroscopy**

We first perform hyperspectral imaging over the mid-IR fingerprint region, generating a stack of 3D chemical images for each mid-IR wavenumber. Extracting the value of a certain voxel in the 3D object from all the 3D chemical images in the mid-IR fingerprint region and repeating this procedure over all voxels can generate the raw 3D site-specific mid-IR spectrum. Then, the raw spectrum is normalized by the measured mid-IR laser intensity spectrum and further denoised using a Savitzky-Golay filter[79]. We repeat the above procedures for the voxels of interest selected per depth. To increase the SNR, we average the obtained spectra within the same depth to generate the depth-resolved mid-IR spectra. We further performed baseline correction processing on the mid-IR spectra[60]. For protein secondary structure analysis, we selected the amide I band region (1600 cm$^{-1}$ to 1700 cm$^{-1}$) to perform the deconvolution on the spectra assuming Lorentzian lineshape[46,60]. The abovementioned baseline correction processing and deconvolution-based secondary structure analysis were all performed by a peak analysis application module (Peak Deconvolution, Version 1.90) embedded in the commercial software OriginPro 2021(Version 9.8.0.200).

**Biological samples**

Tau RD P301S FRET Biosensor (TRPFB) cells were used for the designated experiment to examine the seeding activity of fibril fractions extracted from aged (9-month) PS19 P301S mouse brain. The TRPFB cells (CAT#CRL-3275) were obtained from the American Type Culture Collection (ATCC). This cell line has been engineered to report tau seeding activity. Tau seeds introduced into the culture media (CRL-3275) can nucleate the aggregation of the endogenous tau reporter proteins[80]. The TRPFB cells were derived by transducing HEK293T cells with two separate lentivirus constructs encoding tau RD P301S-CFP and tau RD P301S-YFP. Dual-positive cells were identified by fluorescence-activated cell sorting and were cloned and isolated using cloning cylinders[80].

The tau fibril fraction extraction is described as follows. We followed similar protocols in the works by Jiang *et al*[81,82]. The frozen hippocampus and cortex tissues were weighed (100mg-250mg) and put in a thick-wall polycarbonate tube (Beckman Coulter Life Sciences, cat # 362305). Then it was homogenized with TBS buffer (50 mM Tris, pH 8.0, 274 mM NaCl, 5 mM KCl) supplemented with protease and phosphatase inhibitor cocktails (Roche, cat#05892791001 and cat#04906837001)[83]. The homogenate was first centrifuged for 20 minutes (28k rpm, 4 °C). Then the pellet was homogenized with buffer B (10 mM Tris, pH 7.4, 800 mM NaCl, 10% sucrose, 1mM EGTA, 1 mM PMSF). The homogenate was centrifuged again for 20 minutes (22k rpm, 4 °C). Then, the supernatant was aliquot to a new thick-wall polycarbonate tube and incubated with 1% Sarkosyl, rotating in a thermomixer at 37 °C for 1 hour. After the incubation, the fraction was further mixed by the ultracentrifuge for 1 hour (55k rpm, 4 °C). Then the sarkosyl-insoluble pellet was resuspended with 50 µl TE buffer (10 mM Tris, 1mM EDTA, pH 8.0), which is the final tau fibril fraction used for the experiment.

After the tau fibril fraction extraction, the seeding process is described as follows. The TRPFB cells were used to detect tau aggregates capable of propagating pathology[80,84,85]. Fractions from 9-month PS19 P301S tau mouse brain (tau fibril fraction, containing 100ng of tau fibrils) were used as the seeds to induce tau aggregation. The vehicle control was applied with the corresponding TE buffer. The cells are fixed after 24 hours. After fixation, the cells were washed five times with $D_2O$-based PBS. Finally, the fixed cells were immersed in $D_2O$ PBS solutions and sandwiched by two Raman-grade 0.2-mm-thick CaF2 glass pieces.

**Software**
Data were acquired by customized MATLAB code. Data were processed by customized MATLAB code, OriginPro 2021 (Version 9.8.0.200), and Fiji ImageJ (Version:1.53 c).

**Data Availability**
All the data are available upon reasonable request to the corresponding authors (jxcheng@bu.edu (J.X.C.) and jianzhao@knights.ucf.edu (J.Z.)).

**Code Availability**
The image reconstruction codes are available upon reasonable request to the corresponding authors (jxcheng@bu.edu (J.X.C.) and jianzhao@knights.ucf.edu (J.Z.)).

**Acknowledgements**
This work is supported by R35GM136223 and a grant from Daylight Solutions to J.X.C and J.Z. The authors thank Dr. Hongjian He for helpful discussions in protein secondary structure analysis.

**Author contributions**
J.Z. and J.X.C. proposed FBS-IDT. J.Z. developed the FBS-IDT system, performed the experiments, processed the data, and wrote the draft. L.J. and B.W. prepared the biological samples and supported the experiments. A.M. developed the IDT-based reconstruction method and assisted with data processing. Y.H.X. assisted with system development, data acquisition, and data processing. J.B.Z. assisted with data processing. H.B.Z. assisted with the manufacture of the probe-laser array. L.T. supported the IDT-based imaging reconstruction and data processing. J.X.C., B.W., and L.T. supervised the research and revised the manuscript. All authors contributed to the final creation of the manuscript.

**Competing interests**
The authors declare no competing interests.